\documentclass[aps,pra,reprint,superscriptaddress,footinbib,]{revtex4-2}

\usepackage[T1]{fontenc}				
\usepackage[utf8]{inputenc}			
\usepackage[english]{babel}			
\usepackage[protrusion=false]{microtype}	

\usepackage{color} 						
\definecolor{red}{rgb}{1,0,0}				
\definecolor{blue}{rgb}{0,0,1}				
\definecolor{black}{rgb}{0,0,0}				

\usepackage{soul} 
\definecolor{hlyellow}{rgb}{0.95,0.95,0}
\definecolor{hlgreen}{rgb}{0,0.95,0}
\sethlcolor{hlyellow}
\soulregister \ref 7
\soulregister \cite 7
\soulregister \citet 7
\soulregister \footnote 8
\soulregister {\ } 7
\soulregister \figref 7
\soulregister \tabref 7
\soulregister \secref 7
\soulregister \appref 7
\soulregister \refref 7
\soulregister \eqref 7
\soulregister \supmat 7

\usepackage{amsmath} 
\usepackage{amssymb,amsfonts,mathrsfs} 
\usepackage{array} 

\usepackage{graphicx} 

\usepackage{hyperref} 
\definecolor{dullmagenta}{rgb}{0.4,0,0.4} 
\definecolor{darkblue}{rgb}{0,0,0.4}
\definecolor{medblue}{rgb}{0,0,0.6}
\definecolor{lightblue}{rgb}{0,0,0.8}
\hypersetup{
	colorlinks=true, 		
	breaklinks=true,		
	linkcolor=lightblue,
	citecolor=lightblue,
	urlcolor=lightblue,
	filecolor=dullmagenta
}
\usepackage[all]{hypcap} 

\usepackage{fixgreek}
\usepackage{physmaths}
\usepackage{physunits}

\newcommand{\figletter}[1]{\textbf{(\mbox{#1})}}
\newcommand{\figref}[1]{Fig.~\ref{#1}} 
\newcommand{\tabref}[1]{Tab.~\ref{#1}} 
\newcommand{\secref}[1]{Sec.~\ref{#1}} 
\newcommand{\refref}[1]{Ref.~\cite{#1}} 
\newcommand{\eqnref}[1]{Eq.~\eqref{#1}} 

\newcommand{\PHC}{\mbox{P.H.-C.}}
\newcommand{\eSiSiGe}{\ce{^28Si}/SiGe}

\newcommand{\NbTiN}{NbTiN}

\newcommand{\appref}[1]{Appendix \ref{#1}} 

\newcommand{\SupRefSecFab}{\appref{sec:fabrication}} 
\newcommand{\SupRefSecSetup}{\appref{sec:setup}} 
\newcommand{\SupRefSecZeroDetuning}{\appref{sec:zerodetuning}} 
\newcommand{\SupRefSecModel}{\appref{sec:model}} 
\newcommand{\SupRefSecSpinSpin}{\appref{sec:extspinspin}} 
\newcommand{\SupRefSecPhotonNumber}{\appref{sec:photonnumber}} 

\newcommand{\SupRefFigExtDevice}{\figref{fig:extdevice}} 
\newcommand{\SupRefFigDispersiveDetuning}{\figref{fig:dipersivedetuning}} 
\newcommand{\SupRefFigExtVRS}{\figref{fig:extvrs1}} 

\soulregister \SupRefSecModel 7

\begin{document}

\newcommand{\mytitle}
{Coherent spin-spin coupling mediated by virtual microwave photons}
\title{\mytitle}

\author{Patrick \surname{Harvey-Collard}}
\email[Correspondence to: ]{P.Collard@USherbrooke.ca}
\affiliation{QuTech and Kavli Institute of Nanoscience, Delft University of Technology, 2628 CJ Delft, The Netherlands}

\author{Jurgen \surname{Dijkema}}
\affiliation{QuTech and Kavli Institute of Nanoscience, Delft University of Technology, 2628 CJ Delft, The Netherlands}

\author{Guoji \surname{Zheng}}
\affiliation{QuTech and Kavli Institute of Nanoscience, Delft University of Technology, 2628 CJ Delft, The Netherlands}

\author{Amir~\surname{Sammak}}
\affiliation{QuTech and Netherlands Organization for Applied Scientific Research (TNO), 2628 CJ Delft, The Netherlands}

\author{Giordano \surname{Scappucci}}
\affiliation{QuTech and Kavli Institute of Nanoscience, Delft University of Technology, 2628 CJ Delft, The Netherlands}

\author{Lieven~M.~K. \surname{Vandersypen}}
\email[Correspondence to: ]{L.M.K.Vandersypen@tudelft.nl}
\affiliation{QuTech and Kavli Institute of Nanoscience, Delft University of Technology, 2628 CJ Delft, The Netherlands}

\date{April 8, 2022}

\begin{abstract}
We report the coherent coupling of two electron spins at a distance via virtual microwave photons. Each spin is trapped in a silicon double quantum dot at either end of a superconducting resonator, achieving spin-photon couplings up to around $g_s/2\pi = 40 \ \text{MHz}$. As the two spins are brought into resonance with each other, but detuned from the photons, an avoided crossing larger than the spin linewidths is observed with an exchange splitting around $2J/2\pi = 20 \ \text{MHz}$. In addition, photon-number states are resolved from the shift $2\chi_s/2\pi = -13 \ \text{MHz}$ that they induce on the spin frequency. 
These observations demonstrate that we reach the strong dispersive regime of circuit quantum electrodynamics with spins. Achieving spin-spin coupling without real photons is essential to long-range two-qubit gates between spin qubits and scalable networks of spin qubits on a chip.
\end{abstract}

\maketitle


\section{Introduction} 

There is tremendous interest in the realization of quantum computers, and architectures based on solid-state devices offer significant advantages to achieve this goal.
Circuit quantum electrodynamics (QED) leverages high-quality-factor superconducting resonators at cryogenic temperatures to enable the coupling and readout of superconducting qubits \cite{blais2004,wallraff2004,majer2007}.
Meanwhile, spin qubits in gate-defined semiconductor quantum dots (QDs) are also promising for quantum computing \cite{loss1998,vandersypen2019}, having achieved high-fidelity quantum operations, long coherence and relaxation times, and operation above 1 kelvin. Spin qubits in silicon could eventually leverage the advanced manufacturing capabilities of the microelectronics industry, which is a compelling argument towards their development.

Significant research efforts have been dedicated to bringing the benefits of circuit QED to the platform of spin qubits \cite{childress2004,  taylor2006,burkard2006}, in order to leverage different characteristics of each system. Experiments have first incoherently coupled photons to QD charges or spins in various materials \cite{petersson2012,  viennot2015,  frey2012,  xu2020a, deng2015}.
Following multiple theory proposals \cite{trif2008,  cottet2010,hu2012a,  beaudoin2016a,srinivasa2016a,harvey2018a,  warren2019a,benito2019c}, 
experiments using spin-charge hybridization have reached the strong coupling regime of circuit QED with single electron spins in silicon \cite{samkharadze2018,  mi2018b}, with multispin qubits in GaAs \cite{landig2018}, and with carbon nanotubes \cite{cubaynes2019a} (albeit not resonantly). 
Concerning distant interactions, many works have demonstrated the coupling of combinations of gallium-arsenide qubits and transmons qubits \cite{woerkom2018,landig2019b,scarlino2019a,  wang2021}.
Recently, resonant spin-spin-resonator coupling has been demonstrated in silicon \cite{borjans2020}. However, the resonant spin-spin-resonator regime does not allow a straightforward two-qubit gate \cite{blais2021}.

Despite these groundbreaking realizations, some important hallmarks of circuit QED experiments have remained elusive for spins, in part due to the insufficient spin-resonator interaction strength, in combination with decoherence and fabrication challenges. These hallmarks include dispersive interaction between two spins mediated by virtual resonator photons \cite{majer2007} and photon-number-dependent spin dispersive shifts \cite{schuster2007}, both requiring a higher level of interaction-to-decoherence ratio than previously achieved \cite{borjans2020}. The former is required for most two-qubit gate schemes and arguably represents the next frontier of the field, while the latter enables higher signal with dispersive readout, and photon state measurement \cite{johnson2010} and universal control \cite{krastanov2015,blais2021}.

In this work, we overcome previous challenges and demonstrate both spin-spin interaction mediated by virtual photons and photon-number-dependent spin dispersive shifts using single spins at either end of a high-impedance superconducting resonator. Each single spin is trapped in a double quantum dot (DQD) formed in a {\eSiSiGe} heterostructure, and tunable spin-charge hybridization is enabled by a micromagnet. 
We first reach the resonant strong spin-photon-spin coupling regime; then, we bring both spins in resonance with each other but detuned from the resonator photons, and we observe a spin-spin avoided crossing showing coherent remote interaction. This differs from previously reported work where the virtual coupling could not be achieved \cite{borjans2020}. Finally, we resolve the photon-number states from the discrete shifts they induce in the spin transition frequency.

\section{Methods}

\begin{figure*}[tbp]
   \centering
   \includegraphics{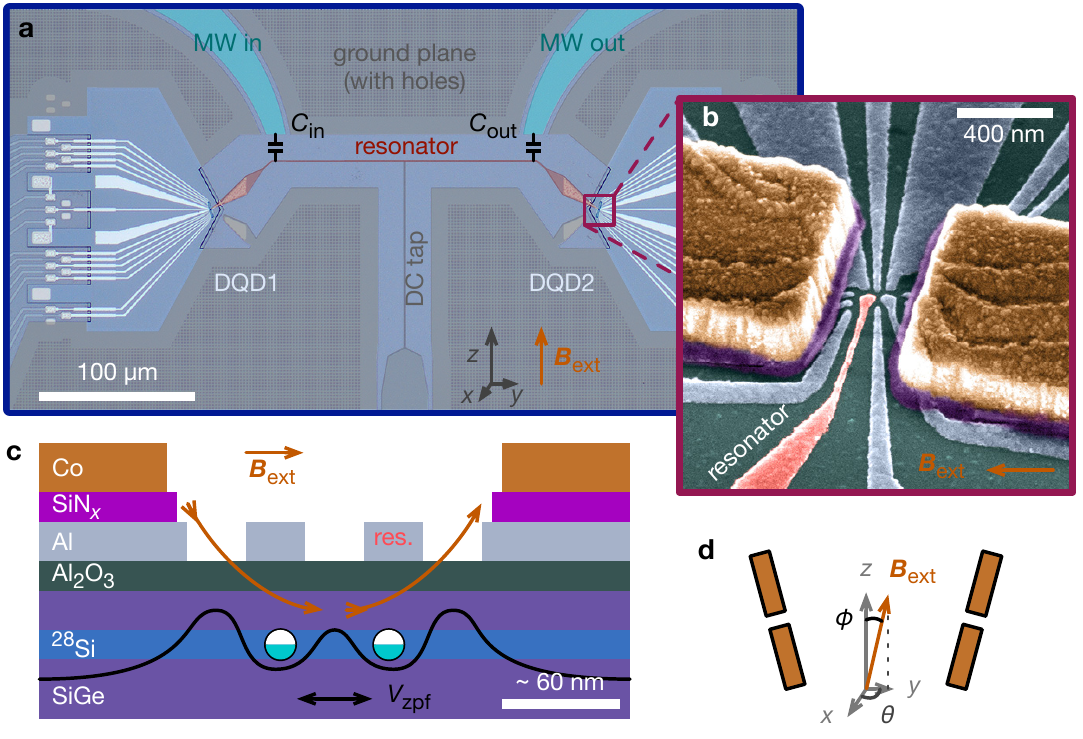}
   \caption{{Spin-spin coupling device.}    \figletter{a}~Optical image of the resonator and QD areas with colored overlays. The ground plane, resonator, microwave (MW) in and out ports, and DC tap are thin superconducting {\NbTiN}. The substantial kinetic inductance and resonator impedance result in a $6.916$-GHz resonator that is only $250 \um$ long.    \figletter{b}~Colorized angled scanning electron microscope image of a (nominally identical) DQD gate structure and micromagnets. The {\NbTiN} resonator end is contacted with an {Al} plunger electrode of the DQD, which is biased through the DC tap. (Image rotated $90 \deg$.)    \figletter{c}~Schematic of the DQD heterostructure and gate stack using the same color scheme as in panel (b). A photon couples to an electron spin through spin-orbit (micromagnet) and orbital-photon (voltage zero-point fluctuations $V_\text{zpf}$) interactions.       \figletter{d}~Micromagnets are tilted $\pm 15 \deg$, allowing one to fine-tune each spin's Zeeman energy through the external magnetic field angle $\phi$. Here, $\vct B_\text{ext} = (B_r, \theta=-90\deg, \phi)$.}
   \label{fig:device}
\end{figure*}

The device is shown in \figref{fig:device}.
A 5-to-7-nm-thin film of {\NbTiN} is deposited on the surface of a {\eSiSiGe} heterostructure and patterned to form the superconducting resonator, ground planes, gate filters \cite{mi2017a,harvey-collard2020a}, and gate fan-out lines (\SupRefFigExtDevice{}). The sheet kinetic inductance (around $140 \pHpsq$), the narrow width of the resonator center conductor ($170 \nm$), and the retracted ground planes combine into a high effective resonator impedance, $Z_r \approx 3.0 \kohm$ \cite{samkharadze2016,harvey-collard2020a}; this results in a fixed coupling $g_c/2\pi = 192 \MHz$ to the DQD charge degree of freedom (``charge qubit'') at a resonator frequency of $\omega_r/2\pi = 6.916 \GHz$ for the half-wave mode (\figref{fig:device}a). Achieving a large $g_c \propto \alpha_c V_\text{zpf}$ (approximately 5 times larger than \refref{borjans2020}) is a result mostly of the increased voltage zero-point fluctuations $V_\text{zpf} \propto \omega_r\sqrt{Z_r}$ of the high-impedance resonator, and of the $(0,1){-}(1,0)$ interdot transition lever arm $\alpha_c$ \cite{beaudoin2016a}.
A combination of loss mitigation strategies (\SupRefSecFab{}) results in an undercoupled resonator with a linewidth $\kappa_r/2\pi = 1.8 \MHz$, limited by resistive or dielectric losses near the DQDs.
Each end of the resonator terminates as one of the dot plunger gates and is biased through a DC tap. The resonator is only $250 \um$ long, which is a consequence of the substantial kinetic inductance that translates to a high effective magnetic permeability \cite{harvey-collard2020a}.
The DQD potential is shaped by applying suitable voltages to surface gate electrodes, as shown in \figref{fig:device}b-c. 
Cobalt micromagnets provide a transverse magnetic field difference $\Delta B_\perp = 42 \mT$ between the two dots while minimizing magnitude differences $\Delta B_\parallel$ ($\Delta B_\perp$ is taken as constant, see \SupRefSecModel{} for details). 
Unlike $g_c$, which is, to a large extent, fixed by the device structure, the spin-photon coupling $g_s/2\pi$ at zero charge detuning approximatively scales as
\ma{
	g_s \approx g_c \frac{g_\text{e}\mu_\text{B}\Delta B_\perp}{4(2t_c-\hbar\omega_r)}
}
for $2t_c \gg \{\hbar\omega_s, \hbar\omega_r\}$, and it is therefore tunable via the DQD tunnel splitting $2t_c$ \cite{samkharadze2018,mi2018b}. Here, $g_\text{e} = 2$ is the electron Land\'{e} $g$-factor, $\mu_\text{B}$ is the Bohr magneton, and $\omega_s/2\pi$ is the spin transition frequency. In any case, $g_s \leq g_c$.
These device characteristics combine to enable values of $g_s/2\pi$ up to around $40 \MHz$ in this work. 
The micromagnets are tilted $\pm 15 \deg$ relative to the vertical direction (\figref{fig:device}d), allowing one to fine-tune each spin's Zeeman energy by rotating the in-plane external magnetic field of magnitude $B_r = \abs{\vct B_\text{ext}}$ by an angle $\phi$ \cite{borjans2020} using a vector magnet.
Device fabrication and experimental setup details are given in \SupRefSecFab{} and \SupRefSecSetup{}, respectively.

\section{Results}

First, the system is tuned to the regime where both spins are resonant with the photons \cite{borjans2020}, which happens at a field angle of $\phi = 10.5 \deg$ due to a $6.5$-mT difference in the micromagnet fields at $\phi = 0 \deg$. 
\begin{figure*}[tbp]
   \centering
   \includegraphics[width=\textwidth]{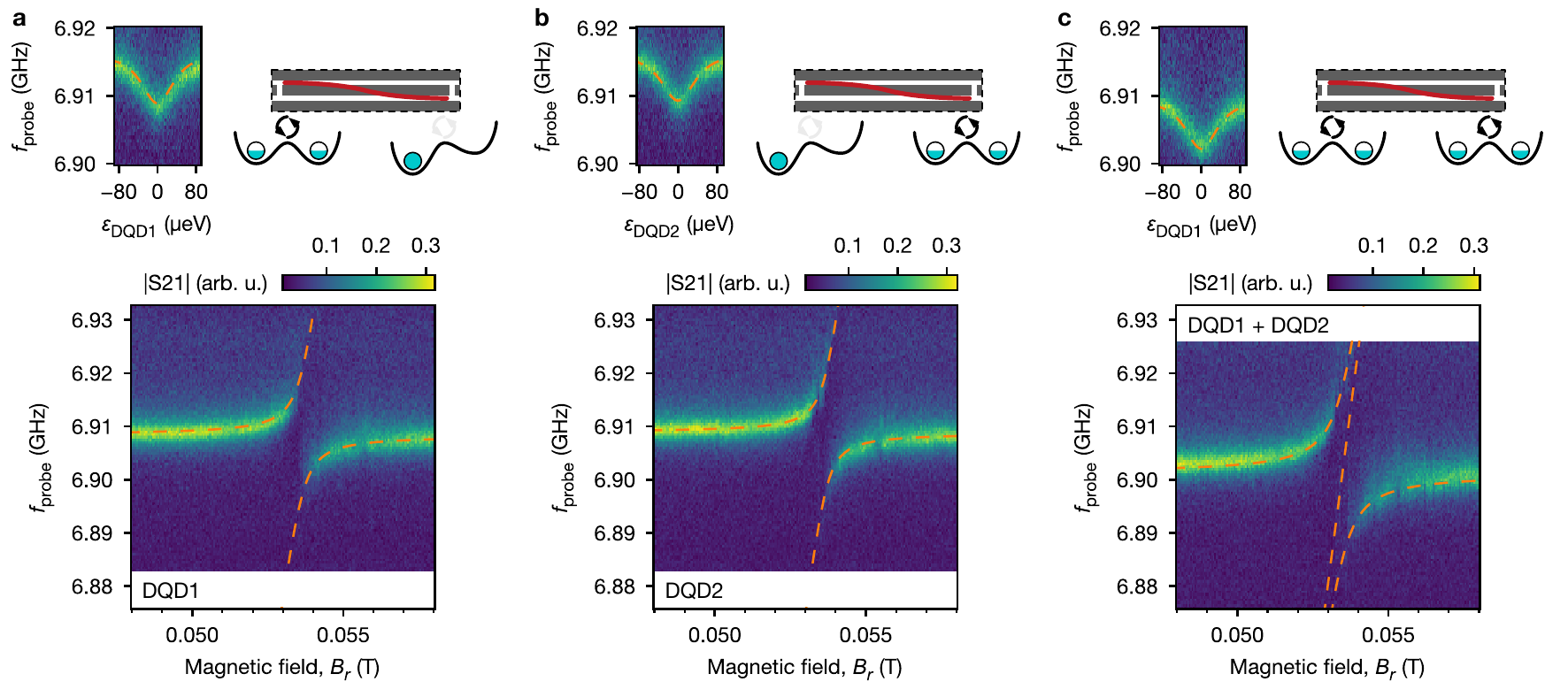} 
   \caption{{Resonant spin-photon-spin coupling.}     \figletter{a-b}~Spin-photon vacuum Rabi splitting of DQD1 (a) and DQD2 (b) for tunnel splittings $(2t_c^1, 2t_c^2)/h = (13.2, 13.7) \GHz$ and a field angle of $\phi = 10.5 \deg$. When the gate voltage is set on the zero-charge-detuning point, which corresponds to the middle of the $(0,1){-}(1,0)$ interdot charge transition, the charge-photon interaction $g_c/2\pi = 192 \MHz$ between the DQD charge and the resonator dispersively shifts the resonator frequency (insets) and enables the spin-photon interaction through the artificial spin-orbit interaction. The dashed lines are the Hamiltonian model transitions. The probe power is $-117.5 \dBm$.     \figletter{c}~When both dots interact simultaneously with the resonator, the dispersive shifts are additive, and the vacuum Rabi splitting is enhanced by a factor of approximately $\sqrt{2}$, an effect of the coherent collective spin-photon interaction. Model transitions are adjusted to the individual interaction data and used to predict the simultaneous interaction data.
   }
   \label{fig:resonant}
\end{figure*}
In the main panels of Figs.~\ref{fig:resonant}a and \ref{fig:resonant}b, DQD1 and DQD2 are set separately to zero charge detuning, $\eps = 0 \ueV$, allowing each spin to interact with the photons while the other is decoupled. For what follows, it is useful to keep in mind that the effective charge-photon coupling (and therefore the effective spin-photon coupling) can be switched off simply by biasing the DQD charge detuning ($\epsilon \gg t_c$) such that the electron is confined to one dot. 
Sweeping $B_r$ results in a vacuum Rabi splitting measurement for DQD1 and DQD2. 
The device used in this experiment has slow drift of the $(0,1){-}(1,0)$ interdot charge transitions. Hence, for every magnetic field setting ($B_r$ or $\phi$), an automated fitting procedure is used to extract a data cut along $\eps = 0 \ueV$ and reconstruct the two-dimensional data. This procedure ensures that the measurements are protected against long-term drift, and it is further detailed in \SupRefSecZeroDetuning{}.
A series of measurements is used to successively constrain and extract system parameters, and the calculated transition frequencies are then plotted over the measured data. The model parameters include the resonator frequency $\omega_r/2\pi$, the charge-photon coupling $g_c^i/2\pi$, and the DQD charge detuning and tunnel coupling $\eps^i/h$ and $t_c^i/h$ (respectively). Here, $h$ is the Planck constant, and $i$ is an index identifying the DQD. Full details about the model are given in \SupRefSecModel{}. Experimentally, we observe that there is a high level of symmetry between the two DQDs, and unless mentioned explicitly, we omit the DQD index and take the dot parameters to be the same. 
From the vacuum Rabi splitting measurements of \figref{fig:resonant}a and \figref{fig:resonant}b, we extract a spin-photon coupling of $(g_s^1, g_s^2)/2\pi = (11.8, 11.0) \MHz$ for DQD1 and DQD2, respectively, with tunnel splittings $(2t_c^1, 2t_c^2)/h = (13.2, 13.7) \GHz$. Both DQDs achieve the strong spin-photon coupling regime, i.e., $g_s > \{\kappa'_r, \varGamma_s\}$, where $\varGamma_s \leq 6 \MHz$ is the spin linewidth. 
When both DQDs are set to zero charge detuning simultaneously (\figref{fig:resonant}c), both spins interact with the resonator, yielding a larger resonator dispersive shift due to the additive effects of the two DQDs, as well as an enhanced vacuum Rabi splitting $2g_s^{12}/2\pi = 32.3 \MHz$ from the two-spin ensemble. The enlarged splitting matches well the predicted $2\sqrt{(g_{s}^1)^2 + (g_{s}^2)^2}/2\pi$ value, very close to a factor $\sqrt{2}$ larger than for single spins in this case. The Hamiltonian model transitions, which are calibrated solely on the one-at-a-time interaction data, predict very well the outcome of the simultaneous interaction. The $\sqrt{2}$ enhancement indicates simultaneous, coherent, and resonant interaction of both spins with the resonator, as demonstrated in prior work~\cite{borjans2020}.

\begin{figure*}[tbp]
   \centering
   \includegraphics[width=\textwidth]{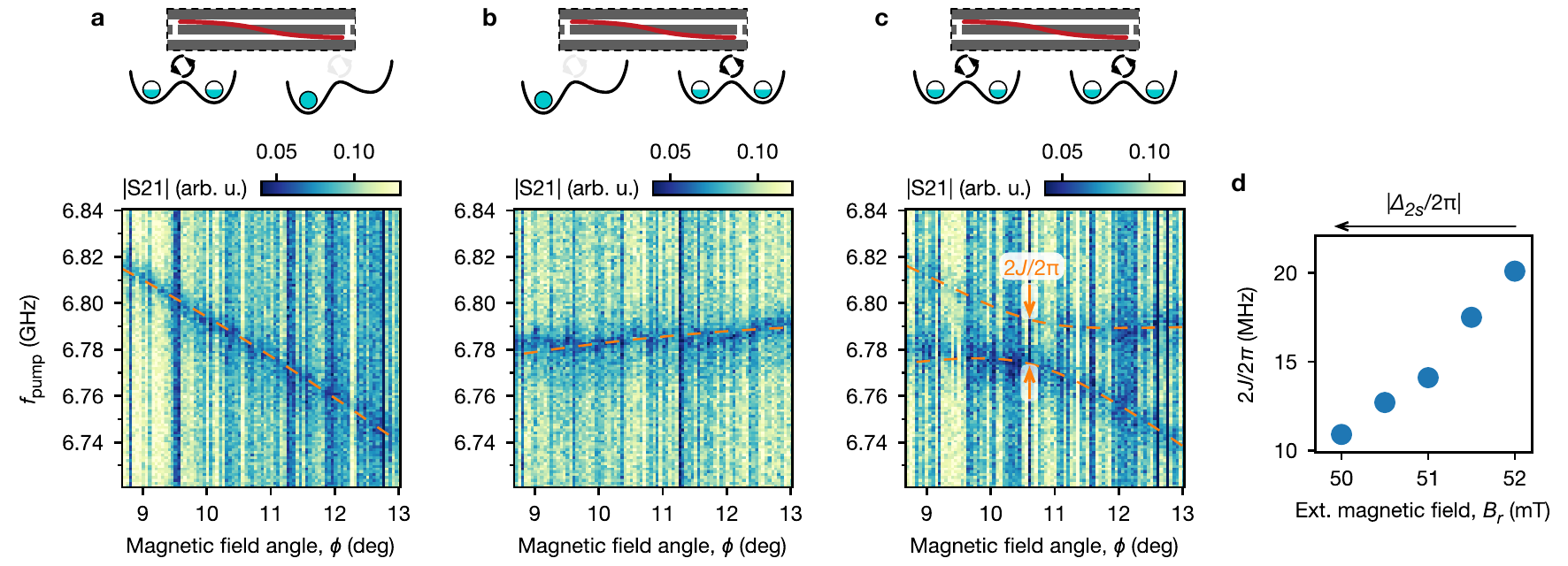} 
   \caption{{Spin-spin coupling via virtual photons.}     \figletter{a-b}~Dependence of the independently interacting spin transition frequencies as a function of the external magnetic field angle $\phi$ at $B_r = 52 \mT$. See \SupRefSecZeroDetuning{} for the reconstruction procedure responsible for the vertical stripes.     \figletter{c}~When both spins interact simultaneously with the resonator, an avoided crossing is observed. The spin states hybridize with a splitting $2J/2\pi = 19.0 \MHz$, larger than any of the transition widths $2\varGamma_s/2\pi \leq (13.2\pm4.2) \MHz$. A dark state is observed along the upper branch, an effect caused by the symmetry of the coherently hybridized spin states.     \figletter{d}~Plot showing how the exchange interaction is reduced when the spin-photon detuning is increased.}
   \label{fig:dispersivephi}
\end{figure*}

To resolve the spin-spin exchange splitting $2J/2\pi$ mediated by virtual photons, a larger spin-photon coupling is needed than in the resonant case (see, e.g., \SupRefFigExtVRS{}). Given a fixed ratio $g_s/\varDelta_s$, where $\varDelta_s = \omega_s - \omega_r$ is the spin-photon detuning, and given that $J \approx (g_s)^2/\varDelta_s$ in the dispersive regime, increasing $g_s$ should allow $J$ to become larger than the spin linewidth.
The spin-photon coupling strength is increased by reducing the DQD tunnel splitting to $2t_c/h \approx 8.8 \GHz$, yielding $g_s/2\pi \approx 33 \MHz$. 
Insight into how the two spin states hybridize is gained by measuring the spin transition frequencies as a function of the external magnetic field angle $\phi$ using two-tone spectroscopy. A pump tone at frequency $f_\text{pump}$ is sent down a gate line to each DQD to generate an excited spin-up population, while the transmission coefficient {S21} is probed at a fixed frequency $f_\text{probe}$ set to the dispersively shifted resonator frequency for each DQD at zero charge detuning (e.g., as in the insets of \figref{fig:resonant}).
Line cuts along the charge zero-detuning axis of data like those of \SupRefFigDispersiveDetuning{} are assembled in a two-dimensional diagram, resulting in \figref{fig:dispersivephi}.
The spin transition of the independently interacting DQD1 and DQD2 is visible as a dip in the {S21} signal magnitude. The slope in the spin transition frequency as a function of $\phi$ is caused by the relative angle of the field and each micromagnet, which allows one to tune each spin's transition energy, as explained earlier. The DQD2 slope is smaller than the one of DQD1 because $\phi \in [9, 13] \deg$ is almost aligned with the DQD2 micromagnet angle $15 \deg$ and farther from the one of DQD1. When both spins interact simultaneously (\figref{fig:dispersivephi}c), an avoided crossing is observed, while the upper transition becomes dark close to spin-spin resonance \cite{majer2007}. With $(2t_c^1, 2t_c^2)/h = (8.82, 8.80) \GHz$ and $B_r = 52 \mT$, we extract $(g_s^1, g_s^2)/2\pi = (32.4, 32.7) \MHz$, $2J/2\pi = 19.0 \MHz$, and $\varDelta_{2s}/2\pi = -79 \MHz$ (here, $\varDelta_{2s}$ is for two spins since each DQD contributes an additive charge dispersive shift $\chi_c$ to $\w_r$, as explained in \SupRefSecModel{}). In \figref{fig:dispersivephi}d and \SupRefSecSpinSpin{}, additional spin-spin hybridization results show that the exchange interaction is reduced when either $\abs{\varDelta_s}$ is increased or $g_s$ is decreased, as expected. The full width at half minimum of the dip, $2\varGamma_s/2\pi$, is $2\varGamma_s^-/2\pi = (11.7 \pm 3.6) \MHz$ for the two-spin lower branch on resonance (the upper branch's visibility is too low for a reliable fit). While this is the most relevant linewidth, for completeness, we also report $(2\varGamma_s^1, 2\varGamma_s^2)/2\pi = (10.4\pm3.0, 13.0\pm3.2) \MHz$ for the individual spin-interaction case on resonance ($\phi = 10.5 \deg$), and $(2\varGamma_s^1, 2\varGamma_s^2)/2\pi = (8.8\pm4.1, 13.2\pm4.2) \MHz$ for the simultaneous interaction away from resonance ($\phi = 13 \deg$). Since $2J > (2\varGamma_s^1 + 2\varGamma_s^2)/2$, the two spins are coherently hybridized through virtual photons, achieving a long-standing goal for the field.
Arguably, the data presented in this figure are not very deep into the dispersive regime. In \SupRefSecSpinSpin{}, we present additional data at larger $\varDelta_s$ where $2J$ is still larger than $2\varGamma_s$.
The ratio of interaction strength to decoherence should be sufficient to enable two-qubit gates \cite{srinivasa2016a,harvey2018a,benito2019c,warren2019a} in future experiments. 

\begin{figure*}[tbp]
   \centering
   \includegraphics[width=\textwidth]{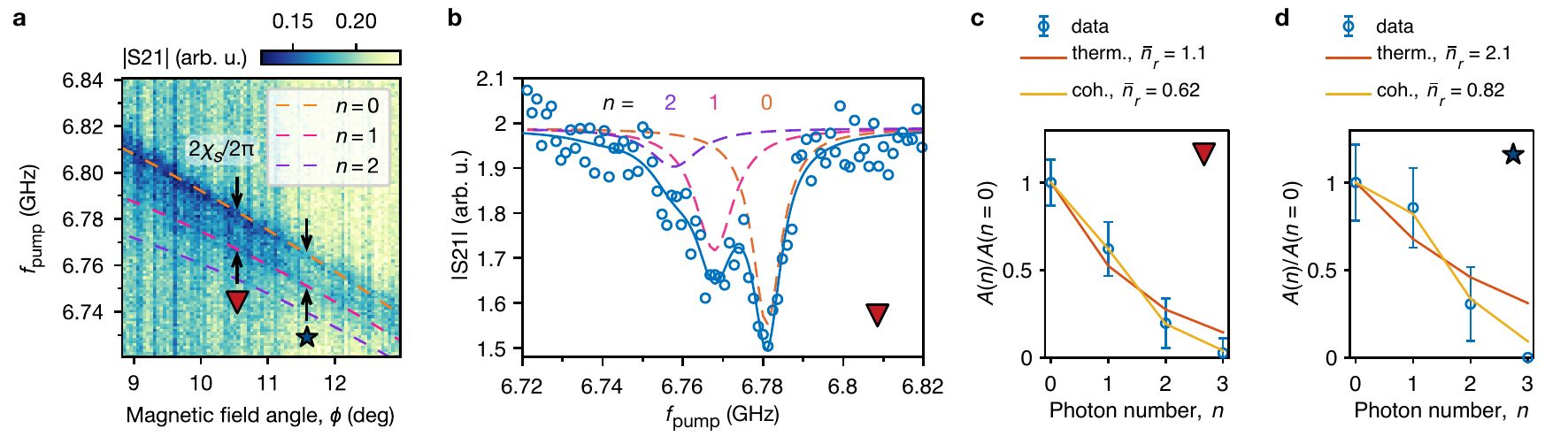} 
   \caption{{Photon-number-resolved spin dispersive shift.}     \figletter{a}~Spin transition frequency of DQD1 showing an extra dip below the main transition for larger probe power. The dip matches well the prediction from the Hamiltonian model (dashed lines) for $\ket{\downarrow,n} \leftrightarrow \ket{\uparrow,n}$ transitions, with the first two transitions visible in this plot.     \figletter{b}~Line cut of the data in panel (a) for $\phi = 10.55 \deg$, and fit to four Lorentzian dips (first three shown with dashed lines). A dispersive shift $2\chi_s/2\pi = -13.1 \pm 2.2 \MHz$ is extracted.     \figletter{c}~Relative area of the dips for $\phi = 10.55 \deg$, and comparison with thermal [$P_\text{thermal}(n) = \bar n_r^n/(\bar n_r+1)^{n+1}$] and coherent [$P_\text{coherent}(n) = \E{-\bar n_r}\bar n_r^n/n!$] photon-number distributions. The error bar represents the $95 \pc$ confidence interval.     \figletter{d}~Relative area for $\phi = 11.6 \deg$. At this angle, the resonator has a smaller $\kappa'_r$, resulting in a larger steady-state photon number. The coherent-state distribution has better agreement.}
   \label{fig:dispersivephotonnumber}
\end{figure*}

The photon-number-dependent dispersive shift, where a single (probe) photon will shift the qubit frequency by more than its linewidth, is also a hallmark of circuit QED \cite{schuster2007}. The shift $2\chi_s/2\pi$ is expected to scale as $\chi_s \approx (g_s)^2/\varDelta_s$ in the dispersive regime, a dependence reminiscent of the qubit-qubit exchange interaction $J$. It therefore seems reasonable to observe both effects if the linewidths are sufficiently narrow. However, while virtual spin-spin coupling requires zero photons and therefore has lower sensitivity to photon losses, this effect requires finite photon population, which exposes the spin to measurement broadening (see the $\varGamma(n)$ formula below).
The photon-number-dependent dispersive shift of DQD1 is shown in \figref{fig:dispersivephotonnumber}.
A larger probe power is used ($-119 \dBm$, $0 \MHz$ detuning) than for the data of \figref{fig:dispersivephi} ($-123 \dBm$, $-1.2 \MHz$ detuning), which populates the resonator with more photons. 
An extra dip appears below the main transition, and its frequency shows good agreement with the prediction of the Hamiltonian model for the $\ket{\downarrow,1} \leftrightarrow \ket{\uparrow,1}$ spinlike transition. 
In \figref{fig:dispersivephotonnumber}b, a line cut from \figref{fig:dispersivephotonnumber}a is extracted, and the dip areas and separations are fit to a sum of Lorentzian dips with widths $2\varGamma(n)/2\pi = 2\gamma_s/2\pi + (n + \bar n_r)\kappa'_r/2\pi$. A value of $2\chi_s/2\pi = -13.1 \pm 2.2 \MHz$ is extracted from the fit (with parameters $\kappa'_r/2\pi = 3.0 \pm 0.2 \MHz$, $\bar n_r = 0.62$), slightly larger than the linewidths $(2\varGamma(0), 2\varGamma(1))/2\pi = (8.6, 11.6) \MHz$. The Hamiltonian model yields $g_s/2\pi = 33.4 \MHz$ and $\varDelta_s/2\pi = -102 \MHz$. The dressed linewidth $\kappa'_r$ accounts for exact experimental conditions at the time of the measurement, including Purcell decay caused by the charge qubit that could otherwise change slightly over time and conditions.
The area under the photon-number dips should be proportional to the probability of each photon number $n$ \cite{schuster2007,gambetta2006}. The relative areas are plotted in \figref{fig:dispersivephotonnumber}c and compared with thermal- and coherent-state distributions. The coherent state with $\bar n_r = 0.62$ shows better agreement. This is consistent with the observation that the areas of the extra photon-number dips are reduced when using lower probe powers. Additional analysis can be found in \SupRefSecPhotonNumber{}.

The ability to resolve quantized photon-number shifts in the qubit spectrum is a feature of the so-called strong dispersive regime of circuit QED, $\chi_s > \{\varGamma_s, \kappa'_r\}$. It enables the preparation and detection of quantum photon states such as number states or cat states \cite{johnson2010}, and therefore paves the way towards bosonic codes \cite{mirrahimi2014,krastanov2015} with spin qubits. Reciprocally, this also entails a shift of the resonator frequency larger than its linewidth, enabling fast and strong qubit readout. Finally, it dramatically highlights the consequences of residual photons on the qubits' dephasing.

\section{Discussion}

The strong backaction of the probe photons on the spin observed here highlights the limits of continuous-wave measurements, and the necessity for future work to include time-domain control or dedicated readout and coupling resonators, as is now standard with superconducting qubits.
Since the probe power needs to be much below $\bar n_r=1$ photons, this entails a low signal-to-noise ratio, especially without a parametric amplifier. Because of slow drift in the DQD interdot transition specific to this device (of the scale of minutes), long averaging times become problematic. However, this is not a fundamental problem for the platform in the future, as spin-qubit devices can be much more stable.

The fit to the formula for $\varGamma(n)$ suggests a fundamental spin linewidth $\gamma_s/2\pi = 3.4 \pm 1.0 \MHz$ when subtracting photon losses ($\kappa'_r/2\pi = 3.0 \pm 0.2 \MHz$) and for $g_s/2\pi = 33.4 \MHz$. This linewidth value should be cited with care given that there are many assumptions involved; however, it compares favorably with literature values \cite{samkharadze2018,mi2018b,landig2018}, and especially when considering the large $g_s$ achieved here.
As with previous work, the spin linewidth seems limited by charge noise coupling in through the artificial spin-orbit interaction.

The large $g_s$ in this experiment is key to achieving coherent spin-photon interaction in the dispersive regime. This would not be possible without the large $g_c$ enabled by the high-impedance resonator, considering that $g_s \leq g_c$ \cite{warren2019a,benito2019c}. Other contributing factors include the engineered spin-orbit interaction through the micromagnet's $\Delta B_\perp$, and state-of-the-art resonator losses despite the increased resonator coupling to its environment \cite{harvey-collard2020a}. Interestingly, the charge-qubit linewidth $\varGamma_c/2\pi \lesssim 60 \MHz$ at $2t_c/h = 12 \GHz$ is not particularly small, suggesting that the good overall performance of the device comes from other factors, like the low $g_s/g_c$ ratio \cite{danjou2019a}, and could be improved further \cite{mi2017}. 
The cooperativity reaches a demonstrated value of $(g_s)^2/\kappa'_r\varGamma_s^1 = 72$ (taking a conservative $\varGamma_s^1/2\pi = 5.2 \MHz$) or a projected $(g_s)^2/\kappa'_r\gamma_s = 109$ (assuming photon-induced broadening could be eliminated).
Given that the exchange splitting is larger than the spin decoherence rates, a two-qubit gate of modest fidelity greater than or close to $75\pc$ \cite{benito2019c} could potentially be achieved with the current device. To fully explore the optimal parameter space, a device with improved long-term stability, larger resonator coupling rates for readout, and tailored gate filters to allow driving signals without attenuation would all be beneficial.

\section{Conclusion}

In summary, we have demonstrated coherent hybridization of two spins mediated by virtual photons, as well as spin dispersive shifts by single photons, both larger than the spin linewidth. These experiments are more challenging than previous demonstrations of circuit QED with spins because they require a larger coupling-to-decoherence ratio (i.e., cooperativity). Admittedly, the cooperativity in this platform is not yet on par with contemporary superconducting qubits. 
The regime of circuit QED achieved here is quite promising for the platform; it could enable two-qubit gates between spin qubits mediated by resonators \cite{warren2019a,benito2019c}, single-shot dispersive spin-qubit readout (without spin-to-charge conversion) \cite{danjou2019a}, bosonic codes through preparation and detection of quantum photon states with spins \cite{johnson2010,krastanov2015,albert2018}, coherent links between dense spin-qubit networks \cite{vandersypen2019}, or quantum simulation with spin QED networks \cite{schuetz2019}. 
For future improvements, we believe that decoupling the spin from the charge noise is a promising path. An increased charge-photon coupling would allow to further detune the charge qubit from the spin and photon while preserving the spin-photon coupling, effectively suppressing the effects of the charge linewidth. This is because $g_s \propto (2t_c-\hbar\omega_r)^{-1}$ while $\varGamma_s \propto (2t_c-\hbar\omega_r)^{-2}$ is suppressed more quickly \cite{warren2019a,benito2019c}. Furthermore, improved longitudinal magnetic gradient symmetry (we measured $\Delta B_\parallel \sim 1 \mT$) could help reduce the spin's noise sensitivity. Finally, improvements to materials and fabrication could help reduce the charge noise itself.

\begin{acknowledgments}
The authors thank T.\ Bonsen and M.\ Russ for helpful insights involving input-output simulations, L.~P.\ Kouwenhoven and his team for access to the {\NbTiN} film deposition, F.\ Alanis Carrasco for assistance with sample fabrication, and other members of the spin-qubit team at QuTech for useful discussions.
\paragraph*{\bf Funding}
This research was funded in part by the European Research Council (ERC Synergy Quantum Computer Lab), the Dutch Ministry for Economic Affairs through the allowance for Topconsortia for Knowledge and Innovation (TKI), and the Netherlands Organization for Scientific Research (NWO/OCW) as part of the Frontiers of Nanoscience (NanoFront) program.
\paragraph*{\bf Author contributions}
L.M.K.V., \PHC\ and G.Z.\ conceived and planned the experiments.
\PHC, J.D.\ and G.Z.\ performed the electrical cryogenic measurements.
\PHC\ fabricated the device.
A.S.\ contributed to sample fabrication.
A.S.\ grew the heterostructure with G.S.’s supervision.
\PHC, J.D.\ and L.M.K.V.\ analyzed the results.
\PHC\ wrote the manuscript with input from all co-authors. 
L.M.K.V.\ supervised the project.
\paragraph*{\bf Competing interests}
The authors declare no competing interests.
\paragraph*{\bf Data availability}
The data reported in this paper are archived online at \url{https://dx.doi.org/10.4121/15015453}.

\end{acknowledgments}

\appendix

\newcommand{\RefFigResonant}{\figref{fig:resonant}} 
\newcommand{\RefFigDipersivePhi}{\figref{fig:dispersivephi}} 
\newcommand{\RefFigDipersivePhotonNumber}{\figref{fig:dispersivephotonnumber}} 


\section{Device fabrication}
\label{sec:fabrication}

\begin{figure}
   \centering
   \includegraphics{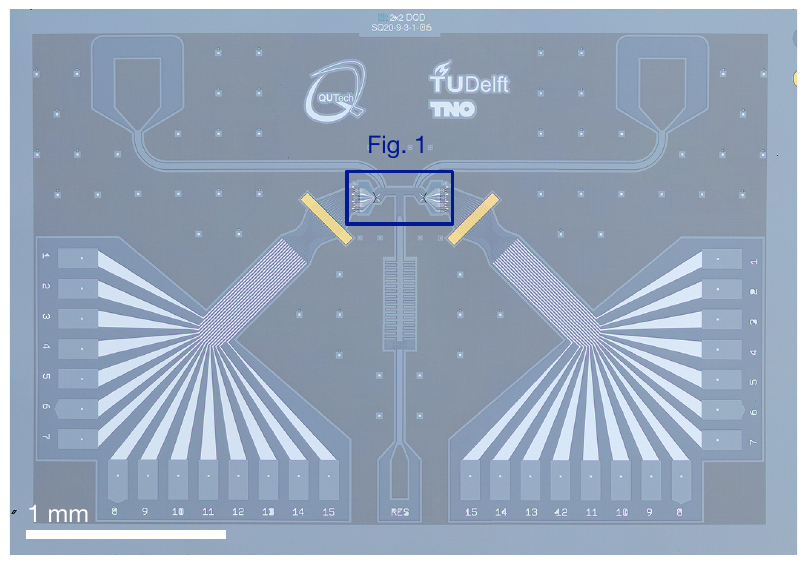} 
   \caption{{Composite optical image of the device chip without wire bonds.}}
   \label{fig:extdevice}
\end{figure}

The 10-nm-thick {\eSiSiGe} quantum-well heterostructure is grown on a $100$-mm {Si} wafer via reduced-pressure chemical vapor deposition. The {SiGe} barrier thickness is $30 \nm$.
Photolithography alignment markers are plasma etched into the surface with a {Cl/HBr} chemistry. 
Doped contacts to the quantum well are formed by \ce{^31P} implantation masked with photolithography and activated with a $700 \degC$ rapid thermal anneal.
The $5{-}7$-nm superconducting {\NbTiN} film is deposited via magnetron sputtering, preceded by a hydrofluoric acid dip and Marangoni drying, and followed by liftoff of the resist-covered quantum dot areas. The sheet inductance is targeted to be around $115 \pHpsq$ (measured at $140 \pHpsq$ for this device). 
The $10$-nm \ce{Al2O3} gate oxide is grown by atomic-layer deposition, followed by wet etching with buffered hydrofluoric acid everywhere except for the resist-covered quantum-dot areas.
Contacts to implants, contacts to the {\NbTiN} film, and electron-beam-lithography alignment markers are patterned with {Ti}/{Pt} evaporation preceded with buffered hydrofluoric acid dip and followed by liftoff.
The wafer is diced into pieces for further electron-beam-lithography steps.
The $25$-nm {Al} gates are deposited via evaporation followed by liftoff.
The {\NbTiN} film is etched via \ce{SF6}/{He} reactive ion etching to define the resonator, inductors, capacitors, and gate lines in a single electron beam lithography step, leaving a $40$-nm step after the etch.
The thin-film capacitor is patterned by first sputtering $30 \nm$ of silicon nitride in a conformal deposition, and then evaporating $5 \nm$ of {Ti} and $100 \nm$ of {Au} in a directional deposition, allowing for a single patterning and liftoff step. The \ce{SiN_z} conformal deposition covers the $40$-nm steps created during the etch of the {\NbTiN} film. 
The micromagnets are patterned by first sputtering $30 \nm$ of silicon nitride in a conformal deposition, and then evaporating $5 \nm$ of {Cr} and $200 \nm$ of {Co} in a directional deposition, allowing for a single patterning and liftoff step.
Pieces are diced into individual $4$-mm-by-$2.8$-mm device chips (\figref{fig:extdevice}) to be wire bonded onto a printed circuit board for cryogenic measurements.

To reduce resonator losses due to resistive currents in the gate structure, the gates are made of {Al}, and they maintain superconductivity up to in-plane magnetic fields of around 0.5 to $0.6 \tesla$, sufficient for our spin-qubit experiments, while the {\NbTiN} structures maintain low losses up to several tesla \cite{samkharadze2016}. To mitigate the microwave losses through the gate fan-out lines \cite{mi2017a}, microwave low-pass filters are patterned on the gate fan-out lines using a combination of nanowire inductors and thin-film capacitors \cite{harvey-collard2020a}. 

\section{Experimental setup}
\label{sec:setup}

\begin{figure*}
   \centering
   \includegraphics{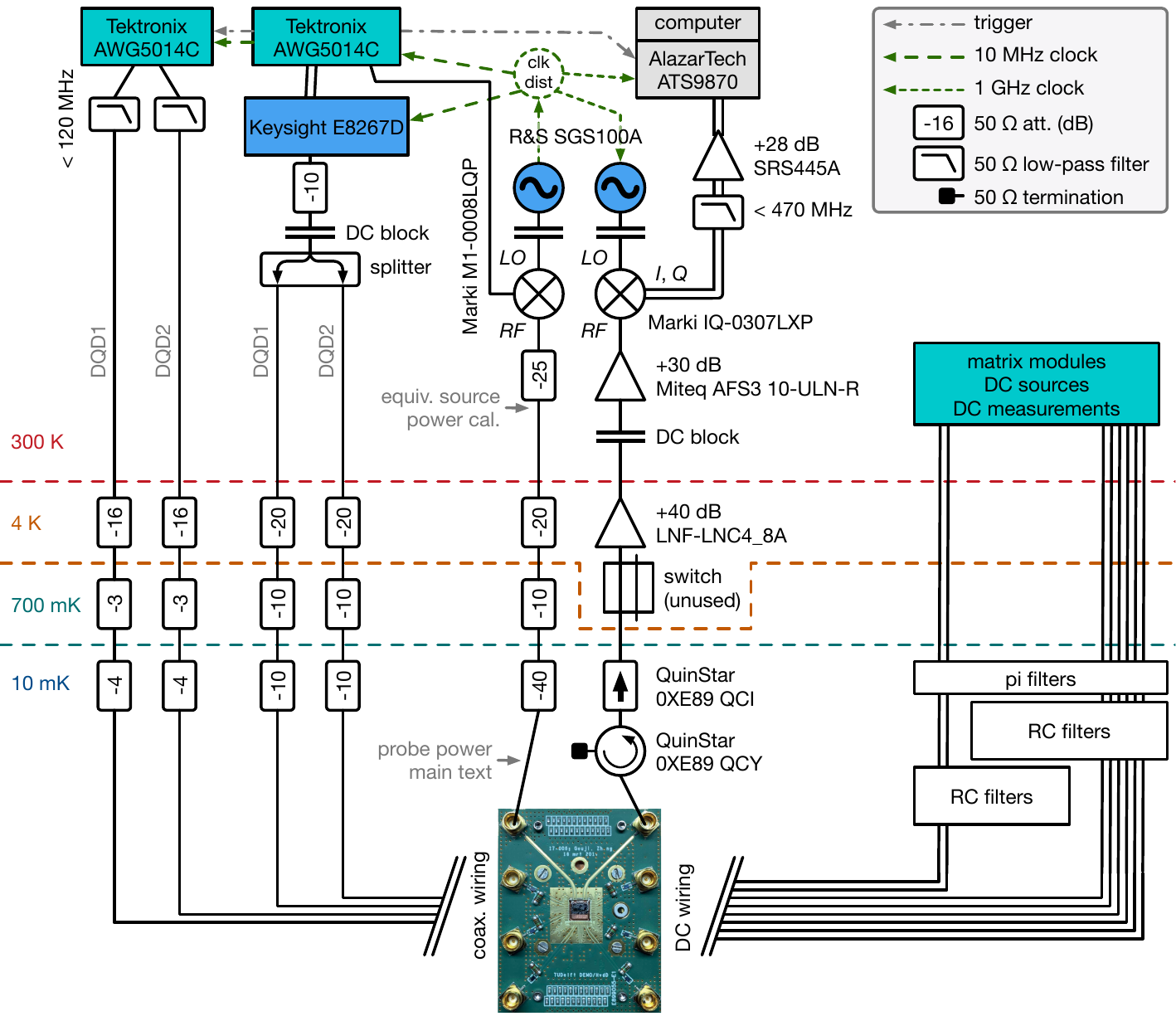} 
   \caption{{Schematic of the measurement setup.}}
   \label{fig:setup}
\end{figure*}

The device is cooled using an Oxford Instruments Triton 400 dilution refrigerator with a base temperature of approximately $8 \mK$. The refrigerator is equipped with a $(6,1,1)$-T vector magnet.
The equipment setup is shown in \figref{fig:setup}.

The resonator probe tone is generated using a mixer to allow for rapid sweeping of the probe frequency. The heterodyne detection is performed with an IQ mixer and at a variable intermediate frequency (IF) in the range $[10, 110] \MHz$. Combined with voltage ramps on the plunger gates or IQ modulation of the pump tone with an arbitrary waveform generator (AWG), this configuration allows for rapid acquisitions of two-dimensional data.

\section{Data reconstruction along zero charge detuning}
\label{sec:zerodetuning}

The device used in this experiment has slow drift of the $(0,1){-}(1,0)$ interdot charge transitions. Hence, for every magnetic field setting ($B_r$ or $\phi$), an automated fitting procedure is used to extract a data cut along $\eps = 0 \ueV$ of a data frame and to reconstruct the two-dimensional data. This procedure ensures that the measurements are protected against long-term drift.

\begin{figure}
   \centering
   \includegraphics{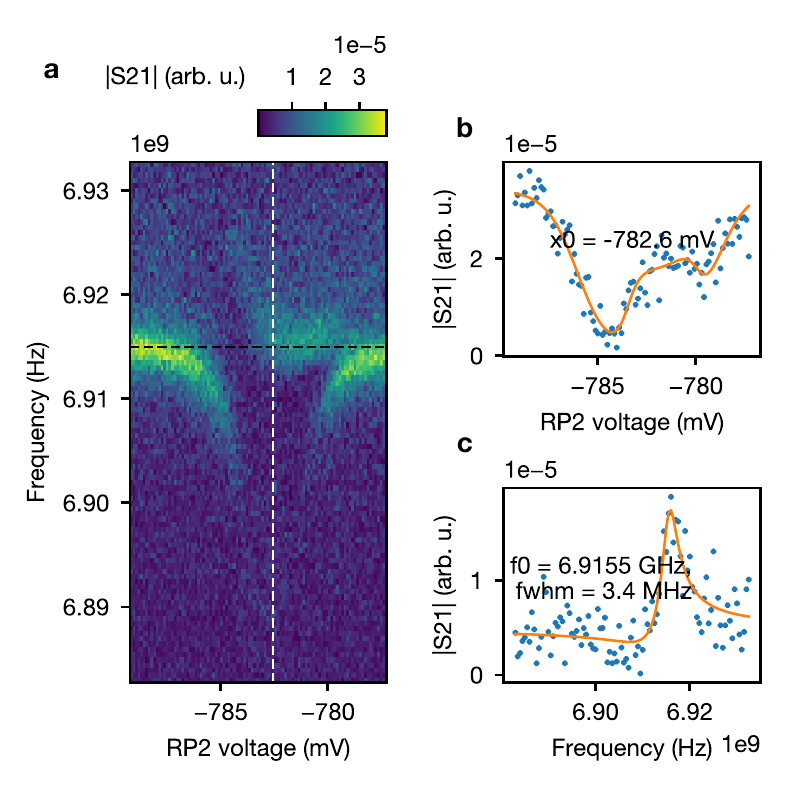} 
   \caption{{Fitting of zero charge detuning.}     \figletter{a}~Example of a single frame of data used in a vacuum Rabi splitting reconstruction.     \figletter{b}~Cut along the black line from panel (a) corresponding to the bare resonator frequency to extract the zero-detuning gate voltage $x_0$. An empiric fitting function is used to find the center of the signal dip due to the dispersive shifts.     \figletter{c}~Cut along the fitted zero-detuning value $x_0$ (white line) and fit to a single Fano resonance.}
   \label{fig:fitcenter}
\end{figure}

Each data frame is acquired in a single digitizer call, as in \figref{fig:fitcenter}a. The fast axis is usually gate voltage, swept with an AWG ramp. The slow axis is usually a pump or a probe frequency, also controlled with AWG I or IQ modulation. The waveforms are repeated with a period of $4 \ms$ (typical) and averaged into a frame consisting of $1000$ to $3000$ repetitions ($4$ to $12 \s$ of cumulative integration time). Data transfer overheads mean that the frames take between 20 and $60 \s$ to acquire and process.

The empirical fitting functions can be found in the source code files used for the data processing. The algorithm works by extracting a line cut at or near the bare resonator frequency, as in \figref{fig:fitcenter}b. The maximum signal is always away from zero charge detuning, and it is symmetrically centered on zero detuning in most cases. For certain values of $B_r$, the spin dispersive shift (positive) compensates the charge dispersive shift (negative) and can lead to features inside the dip, which can look asymmetric. Because these features are always lower in amplitude than the main edges that set the symmetry point, the algorithm is usually robust to this. In some cases, we also improve the fit by accounting for these artifacts, also empirically. The case presented in \figref{fig:fitcenter} is one of those more difficult situations, which works nonetheless (near the spin-photon resonance). The consistency of the whole procedure is validated by visual inspection of the zero-detuning fit results (e.g., by looking at plots like \figref{fig:fitcenter}a one by one). 

For data where both spins interact simultaneously with the resonator, one of the dots is fixed at zero charge detuning while the other dot detuning is rapidly swept, and a similar fitting procedure as described above is applied. This makes the acquisition tolerant of drift for the swept detuning, but the other dot must remain fixed near zero detuning long enough for the automated acquisition to be completed. 
Certain validation procedures can be applied to verify that the fixed dot did not drift after acquisition. For example, the dispersive shifts must add to $(\chi_c^1+\chi_c^2)/2\pi$; away from zero detuning, the shift is less. Conditions are sometimes set to recenter the detuning and reacquire the frame if some conditions are not met. These postvalidation procedures are used best with dispersive spin sensing, but they are harder to implement with vacuum Rabi splitting reconstructions (near spin-photon resonance). They are not always necessary, and are only applied when too much drift is observed. 

Occasional bad data lines are caused by missed zero charge detunings and can easily be seen in the reconstructed data as vertical lines of poor or anomalous signal. Although they are admittedly not pretty, we are confident that they do not impact the validity or interpretation of the results.

The full source data and analysis code is available in the online repository.

\section{Spin-photon Hamiltonian model}
\label{sec:model}

\subsection{Model definition}

The spin-photon system is modeled with the following Hamiltonian to numerically calculate the exact transition energies between eigenstates:
\begin{align}
	H &= \hbar\w_r{a^\dag}a + \sum_{i=1}^2 \ofb{H_i + \hbar g_c^i(a+a^\dag)\tau_z^i} , \label{eq:suph}\\
	H_i &= \frac{1}{2}\ofb{ \eps^i\tau_z^i + 2t_c^i\tau_x^i + \of{\vct h^i + \vct{\Delta h}^i\tau_z^i/2}\cdot\vct\sigma^i } .
	\label{eq:supmodel}
\end{align}
Here, $h$ ($\hbar$) is the (reduced) Planck constant, $a$ is the photon annihilation operator, $\omega_r/2\pi$ is the resonator frequency, $i$ is a DQD index, $g_c^i/2\pi$ is the charge-photon coupling, $\tau_z^i = \ketbra{L^i}{L^i}-\ketbra{R^i}{R^i}$ and $\vct\sigma^i$ are charge and spin Pauli operators (respectively), with $\vct\sigma^i = \sigma_x^i \uvct{x} + \sigma_y^i \uvct{y} + \sigma_z^i \uvct{z}$ a vector of Pauli operators, and $\eps^i/h$ and $2t_c^i/h$ are the DQD charge detuning and tunnel splitting (respectively). In cases where the DQD parameters are the same, or in the context of formulas that describe only one DQD, we omit the DQD index. The external and micromagnet magnetic fields are parametrized using the average and difference field energies $\vct h^i$ and $\vct{\Delta h}^i$ at the left ($L$) and right ($R$) dot positions,
\begin{align}
	\vct h^i/g_\text{e}\mu_\text{B} &= (\vct B^i_L + \vct B^i_R)/2 \\
	&= \vct B_\text{ext} + \vct B^i_\text{um} , \\
	\vct{\Delta h}^i/g_\text{e}\mu_\text{B} &= \vct B^i_L - \vct B^i_R \\
	&= \vct{\Delta B}_\text{um}^i ,
\end{align}
with $g_\text{e} = 2$ the electron Landé $g$-factor and $\mu_\text{B}$ the Bohr magneton. Vectors like the external magnetic field $\vct B_\text{ext} = (B_r, \theta=-90\deg, \phi)$ are conveniently expressed in spherical coordinates, with $\phi$ the polar angle. The micromagnet average field is modeled with the empirical formula
\begin{align}
	\vct B_\text{um}^i = 
	(B_\text{um0}^i + \chi_\text{um}^i(\vct B_\text{ext}-\vct B_\text{ext0}^i) \cdot \uvct u_\text{um}^i) \uvct u_\text{um}^i
\end{align}
to account for the susceptibility ($\chi_\text{um}^1 \approx 0.67$, $\chi_\text{um}^2 \approx 0.63$), and with $\uvct u_\text{um}^1 = (1, 90\deg, 15\deg)$ and $\uvct u_\text{um}^2 = (1, 270\deg, 15\deg)$. Empirical parameters are $B_\text{um0}^1 = 152.6 \mT$, $B_\text{um0}^2 = 146.9 \mT$, and $\vct B_\text{ext0}^i = (B_\text{r0}=-20\mT, -90\deg, \phi) \parallel \vct B_\text{ext}$. However, the micromagnet difference field is taken as constant because of practical difficulties in measuring its response, 
\begin{align}
	\vct{\Delta B}_\text{um}^i = (\Delta B_\perp, 180\deg, 90\deg) = -\Delta B_\perp \uvct x ,
\end{align}
with $\Delta B_\perp = 42 \mT$. We have not noticed discrepancies in $g_s$ (through $\chi_s$ or $J$) that could be specifically attributed to this approximation.
An effect that is left out is the fact that the Zeeman energy is not the same in the left and right dots. Experimentally, we observe $\Delta B_\parallel \sim 1 \mT$ by tracking the spin-photon resonance condition versus DQD charge detuning. This is responsible for some asymmetry in plots like the one of \figref{fig:fitcenter}; however, we find that taking this into account is not necessary to model the energy levels at zero detuning. 
The parametrization described in this paragraph is sufficient to capture the magnitude and angular dependence of the spin Zeeman energies over the range of interest. 

\begin{figure*}[tbp]
   \centering
   \includegraphics{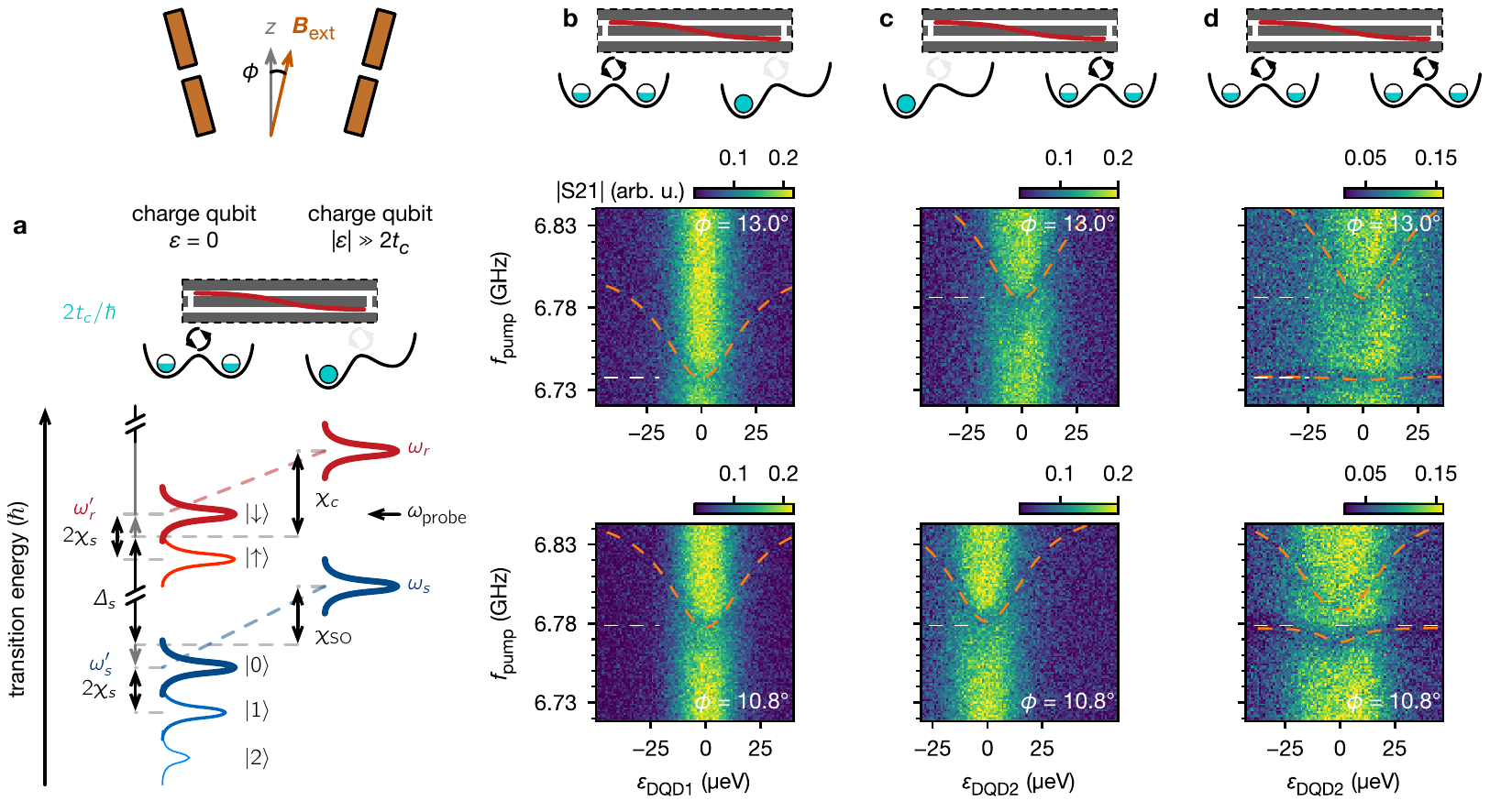} 
   \caption{{Dispersive regime transitions, and spin-spin interaction versus DQD detuning.}     \figletter{a}~Schematic representation of the transition energy shifts in the dispersive regime. See \appref{sec:model} for details.     \figletter{b-c}~Measurement of spin transitions for DQD1 (b) and DQD2 (c) as a function of the charge detuning $\epsilon$ with dispersive readout, with $2t_c/h \approx 8.7 \GHz$ and $B_r = 52 \mT$. A pump tone is sent down the gate lines to excite the spins, resulting in a signal dip when the pump and spin frequencies match. Here, $(f_\text{probe}^1, f_\text{probe}^2, f_\text{probe}^{12}) \approx (6.9038, 6.9061, 6.8981) \GHz$ for $\phi = 10.8 \deg$. The spin transition frequency at $\epsilon = 0 \ueV$ is marked with a white dashed line for comparison between plots.     \figletter{d}~When the spins are simultaneously interacting with the resonator, the spin states hybridize depending on their energy difference. In the nonresonant case, $\phi = 13.0 \deg$, their energies are minimally perturbed, as can be seen by the model lines and the white guides. In the resonant case, $\phi = 10.8 \deg$, the two spins hybridize due to the exchange interaction $2J/2\pi = 20.2 \MHz$ mediated by virtual resonator photons. The upper state is hardly visible because its symmetry makes it dark to the resonator probe, which is expected for a coherent spin-spin interaction.}
   \label{fig:dipersivedetuning}
\end{figure*}

Some derived, effective, spin-photon quantities (such as $g_s$, $\varDelta_s = \w_s-\w_r$, $\chi_s$, $J$) are used by analogy to their idealized versions, without the charge degree of freedom, with, for example, a Tavis-Cummings Hamiltonian \cite{blais2021}
\begin{align}
	\begin{split}
	H_\text{TC} &=  \hbar\w_r{a^\dag}a \\
	&\phantom{= \ }
	+ \sum_{i=1}^2 \ofb{ \frac{\hbar}{2}\w_s^i \sigma_z^i + \hbar g_s^i\of{a^\dagger \sigma_-^i + a \sigma_+^i}} 
	\end{split} \label{eq:hres}
\end{align}
or a dispersive Hamiltonian \cite{blais2021}
\begin{align}
	\begin{split}
	H_\text{disp} &=  \hbar\w_r'{a^\dag}a + \sum_{i=1}^2 \ofb{ \frac{\hbar}{2}\of{\w_s'^i + 2\chi_s^ia^\dagger a}\sigma_z^i } \\
	&\phantom{= \ }
	+ \hbar J\of{\sigma_+^1 \sigma_-^2 + \sigma_-^1 \sigma_+^2} .
	\end{split} \label{eq:hdisp}
\end{align}
Here, $\w_r' = \w_r - \chi_s$ and $\w_s' = \w_s + \chi_s$.
The spin frequency $\w_s/2\pi$ and photon frequency $\w_r/2\pi$ depend strongly on whether each dot interacts with the resonator ($\eps = 0$) or not ($\eps \rightarrow \infty$). The spin frequency is lowered when the interaction is on because of the artificial spin-orbit coupling \cite{benito2019b}, $\w_s \rightarrow \w_s + \chi_\text{SO}$, 
\ma{
	\hbar\chi_\text{SO} &= -\frac{\abs{\vct{\Delta h}}^2}{8\abs{\vct{h}}}
	- \frac{1}{16} \frac{\abs{\vct{\Delta h}}^2}{\of{2t_c-\abs{\vct{h}}}} 
	+ \frac{1}{16} \frac{\abs{\vct{\Delta h}}^2}{\of{2t_c+\abs{\vct{h}}}} 
	< 0 ,
}
for $\vct{\Delta h} \perp \vct{h}$, $2t_c-\abs{\vct{h}} \gg \abs{\vct{\Delta h}}$ and $\abs{\vct{h}} \gg \abs{\vct{\Delta h}}$. 
The photon frequency is lowered by the dispersive interaction with the charge, $\w_r \rightarrow \w_r - \chi_c^i$, for $2t_c > \hbar \w_r$. We find that the charge dispersive shift is well predicted by taking into account the counter-rotating terms (Bloch-Siegert shift),
\ma{
	\chi_c = \frac{(g_c)^2}{2t_c/\hbar-\w_r} + \frac{(g_c)^2}{2t_c/\hbar+\w_r} > 0 .
}
Failing to do so can lead, for example, to an overestimated $g_c/2\pi = 220 \MHz$ instead of $192 \MHz$. A schematic representation of various shifts in the dispersive regime is shown in \figref{fig:dipersivedetuning}a. While the analytical forms are insightful, we use exact numerical values calculated from \eqnref{eq:suph} instead in this work. 
To avoid ambiguities, we define $\varDelta_s^i/2\pi$ as the bare (i.e., theoretical but including $\chi_c$ and $\chi_\text{SO}$) spin-photon detuning for spin $i$ individually interacting with the resonator (see \figref{fig:dipersivedetuning}a), and $\varDelta_{2s}^i/2\pi$ by taking the individually interacting bare spin frequency and the simultaneously interacting bare photon frequency (which we find to be a quite accurate proxy; see $\phi = 13 \deg$ in \figref{fig:dipersivedetuning}).
The dispersive approximations $\chi_s \approx (g_s)^2/\varDelta_s$ and $J \approx g_s^1g_s^2(1/\varDelta_{2s}^1+1/\varDelta_{2s}^2)/2$ are insightful but not quantitatively accurate because of various shifts in the frequencies, rotating wave approximations, dressing by the DQD charge degree of freedom, or violations of the dispersive approximation [requiring $\varDelta_s \gg g_s$ or $\bar n_r \ll n_\text{crit} = (\varDelta_s/2g_s)^2$].
It is not within the scope of this work to derive these quantities from system parameters, as this has been tackled elsewhere \cite{hu2012a,beaudoin2016a,benito2019c,warren2019a}.
Instead, and in spite of Eqs.~\eqref{eq:hres} and \eqref{eq:hdisp}, we define $2g_s/2\pi$ as the vacuum Rabi splitting gap, $2\chi_s/2\pi$ as the shift in the spin frequency induced by one photon, and $2J/2\pi$ as the spin-spin splitting; these quantities are extracted from the data or from Eqs.~\eqref{eq:suph} and \eqref{eq:supmodel}. This avoids the approximation pitfalls mentioned previously. 

\subsection{Parameter extraction from the data}

\begin{figure*}
   \centering
   \includegraphics[width=\textwidth]{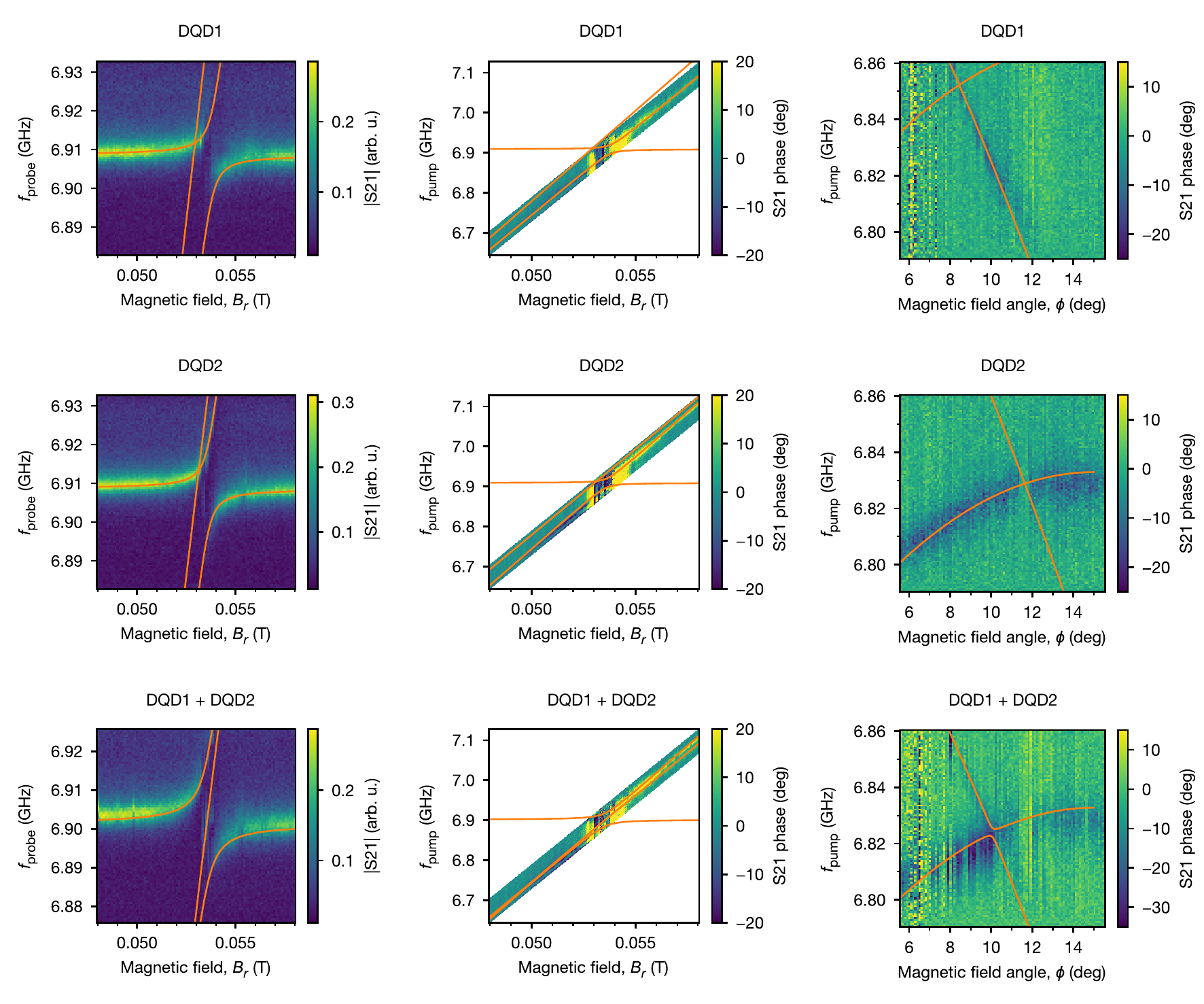} 
   \caption{{Example of data used to extract model parameters.}     For this set, $2t_c/h = 13.5 \GHz$, $B_r = 52 \mT$, and $\phi = 10.5 \deg$. The first column shows vacuum Rabi splitting reconstructions for individually and simultaneously interacting spins. The second column shows dispersive spin sensing (the data near spin-photon resonance are scrambled because of the close proximity of the levels). The third column shows the angular dependence of the spin levels with dispersive spin sensing. The two spinlike and the photonlike transitions are shown together. These data, and more, are used to calibrate the micromagnet parameters for use in the Hamiltonian model \eqnref{eq:supmodel}. For these tunnel splittings and spin-photon detunings, the model predicts $g_s/2\pi = 11.4 \MHz$ and $2J/2\pi = 3.3 \MHz$. Here, $2J/2\pi$ is too small to be resolved. Note that transition linewidths are power broadened.}
   \label{fig:extvrs1}
\end{figure*}

\begin{table*}
   \centering
   \caption{{Summary of model parameters.} The resonator linewidth is a relaxation rate ($\kappa_r=\kappa_1$), while the charge and spin (the ``qubits'') linewidths are dephasing rates, i.e., $\gamma_2=\gamma_1/2+\gamma_\phi$, as is standard in the field \cite{blais2021}. In general, the frequencies and shifts depend strongly on whether one, two, or zero of the DQDs are interacting with the resonator ($\epsilon^i = 0 \ueV$).}
   \begin{tabular}{l l r r r}
	\hline \hline 
	\bf Quantity, symbol & \bf units & \bf DQD1 & \bf DQD2 & \bf resonator \\ \hline
	noninteracting resonator frequency, $\omega_r/2\pi$			&GHz&&&6.916\\
	noninteracting resonator linewidth, $\kappa_r/2\pi$				&MHz&&&1.8\\
	charge-photon coupling, $g_c/2\pi$					&MHz&192&192&\\
	charge-qubit linewidth ($2t_c/h=12\GHz$), $\varGamma_c/2\pi$		&MHz&$\lesssim60$&$\lesssim60$&\\
	transverse magnetic field difference, $\Delta B_\perp$	&mT&42&42&\\
	\hline
	\multicolumn{5}{l}{\bf Resonant coupling, \RefFigResonant{}} \\
	tunnel splitting, $2t_c/h$							&GHz&13.2&13.7&\\
	charge dispersive shift, $\chi_c/2\pi$					&MHz&7.689&7.214&\\
	spin-orbit shift, $\chi_\text{SO}/2\pi$					&MHz&-34.4&-33.6&\\
	spin-photon coupling, $g_s/2\pi$					&MHz&11.8&11.0&\\
	half vacuum Rabi splitting (two spins), $g_s^{12}/2\pi$	&MHz&&&16.2\\ 
	spin linewidth, $\varGamma_s/2\pi$					&MHz&$\leq6$&$\leq6$&\\
	interacting resonator linewidth, $\kappa'_r/2\pi$			&MHz&2.6&2.3&\\
	\hline
	\multicolumn{5}{l}{\bf Dispersive coupling, \RefFigDipersivePhi{}} \\
	tunnel splitting, $2t_c/h$							&GHz&8.82&8.80&\\
	charge dispersive shift, $\chi_c/2\pi$					&MHz&21.46&21.66&\\
	spin-orbit shift, $\chi_\text{SO}/2\pi$					&MHz&-62.3&-62.6&\\
	spin-photon coupling, $g_s/2\pi$					&MHz&32.4&32.7&\\
	spin-photon detuning, $\Delta_s/2\pi$				&MHz&-99&-101&\\
	spin-photon detuning (two-spins), $\Delta_{2s}/2\pi$		&MHz&-79&-79&\\
	spin dispersive shift, $\chi_s/2\pi$					&MHz&-7.8&-7.8&\\
	spin-spin coupling, $J/2\pi$						&MHz&&&9.5\\
	spin linewidth, $\varGamma_s/2\pi$					&MHz&5.2&6.6&\\
	mean photon number, $\bar n_r$					&&&&<0.5\\
	\hline
	\multicolumn{5}{l}{\bf Photon number splitting, \RefFigDipersivePhotonNumber{}, $\phi = 10.55 \deg$} \\
	tunnel splitting, $2t_c/h$							&GHz&8.75&&\\
	charge dispersive shift, $\chi_c/2\pi$					&MHz&22.18&&\\
	spin-orbit shift, $\chi_\text{SO}/2\pi$					&MHz&-63.7&&\\
	spin-photon coupling, $g_s/2\pi$					&MHz&33.4&&\\
	spin-photon detuning, $\Delta_s/2\pi$				&MHz&-102&&\\
	spin dispersive shift (predicted), $\chi_s/2\pi$			&MHz&-8.0&&\\
	spin dispersive shift (measured), $\chi_s/2\pi$			&MHz&-6.6&&\\
	zero-photon-peak linewidth, $\varGamma(0)/2\pi$			&MHz&4.3&&\\
	one-photon-peak linewidth, $\varGamma(1)/2\pi$			&MHz&5.8&&\\
	spin linewidth without photon losses, $\gamma_s/2\pi$	&MHz&3.4&&\\
	interacting resonator linewidth, $\kappa'_r/2\pi$			&MHz&&&3.0\\
	mean photon number, $\bar n_r$					&&&&0.62\\
	\hline \hline
\end{tabular}
   \label{tab:params}
\end{table*}

Here, we describe the procedure to extract the model parameters from the experiment.
Parameters are successively constrained using specific experiments for each DQD one at a time.
First, the resonator bare frequency and bare linewidth are easily extracted from a probe frequency sweep while the dots are in Coulomb blockade. These quantities are later dressed by the interaction with charge and spin. For example, the resonator linewidth is significantly affected by Purcell decay from the charge qubit, and this practically limits the achievable spin-photon coupling and spin measurement sensitivity.
The DQD lever arm is extracted from bias triangles.
The charge-qubit transition frequency $2t_c/h$ is measured with two-tone spectroscopy, and the corresponding resonator dispersive shift (away from spin-photon resonance) then allows one to uniquely calibrate the value of $g_c$.
Next, the micromagnet parameters are extracted by simultaneously adjusting the spin-photon transition frequencies to the experimental values of \figref{fig:extvrs1} (vacuum Rabi splitting, and two-tone spin measurement versus $B_r$ and $\phi$) for each dot independently.
Notably, the size of the spin-photon gap ($2g_s/2\pi$) in a vacuum Rabi splitting measurement is most affected by $\Delta B_\perp$. The micromagnet susceptibility $\chi_\text{um}$ and offset field $B_\text{um0}$ are mainly fixed by the slope of the spin transition frequency versus $B_r$ and the spin-photon resonance condition, respectively. Then, $B_\text{r0}$ is tweaked to get the best simultaneous agreement with both dots and for various angles.

\subsection{Parameter table}

A summary of the main model parameters for the key results is given in \tabref{tab:params}.

\section{Spin transitions as a function of DQD detuning}
\label{sec:dispdetuning}

The DQD tunnel splitting is set to $2t_c/h = 8.7 \GHz$, yielding $g_s/2\pi = 34.2 \MHz$. A pump tone at frequency $f_\text{pump}$ is sent down a gate line to each DQD to generate an excited spin-up population, while the transmission coefficient {S21} is probed at a fixed frequency $f_\text{probe}$ set to the dispersively shifted resonator frequency for each DQD at zero charge detuning (e.g., as in the insets of \RefFigResonant{}).
The results are plotted in \figref{fig:dipersivedetuning} for two values of the magnetic field angle $\phi$, corresponding to off-resonant and resonant spin transition energies. The spin transition frequency is reduced at zero charge detuning compared with the localized dot states because of the spin-charge hybridization shift $\chi_\text{SO}/2\pi$. The signal shows a peak centered on $\epsilon = 0 \ueV$ when the resonator frequency gets pushed down by the charge dispersive shift $\chi_c/2\pi$. When the pump frequency matches the spin transition frequency, the excited spin-state population is increased, and the resonator frequency is pushed down further by the spin dispersive shift $2\chi_s/2\pi$, visible as a dip in the transmission that follows the spin transition energy.
When the two spins are not resonant ($\phi = 13.0 \deg$), each spin can be independently measured while interacting with the resonator one at a time. When the two spins are simultaneously interacting, they can both still be sensed, albeit with an adjusted probe frequency to account for the dual charge dispersive shift, and the spin transitions are mostly unperturbed from their independent values, as seen from the white dashed line serving as a guide.
When the two spins are set to resonance ($\phi = 10.8 \deg$ at this magnetic field), the two states avoid each other when both spins are simultaneously interacting. 
Comparing the simultaneous spin-interaction results in the two cases, we see that the lower state has enhanced visibility while the upper state has a reduced one. This is consistent with the formation of a dark state, an effect that results from the symmetry of the hybridized spin states and is expected in the case of coherent spin-spin interaction \cite{majer2007}.
As in the other cases, model transitions are adjusted to the individual interaction data and used to predict the simultaneous interaction data. The simultaneous interaction is again well predicted by the Hamiltonian model, as can be seen from the orange dashed lines. 
From the model, we extract a minimum separation between the spin states of $2J/2\pi = 20.2 \MHz$, with $B_r = 52 \mT$ and $\varDelta_{2s} = -80 \MHz$.

\section{Extended spin-spin coupling data}
\label{sec:extspinspin}

\begin{figure*}
   \centering
   \includegraphics{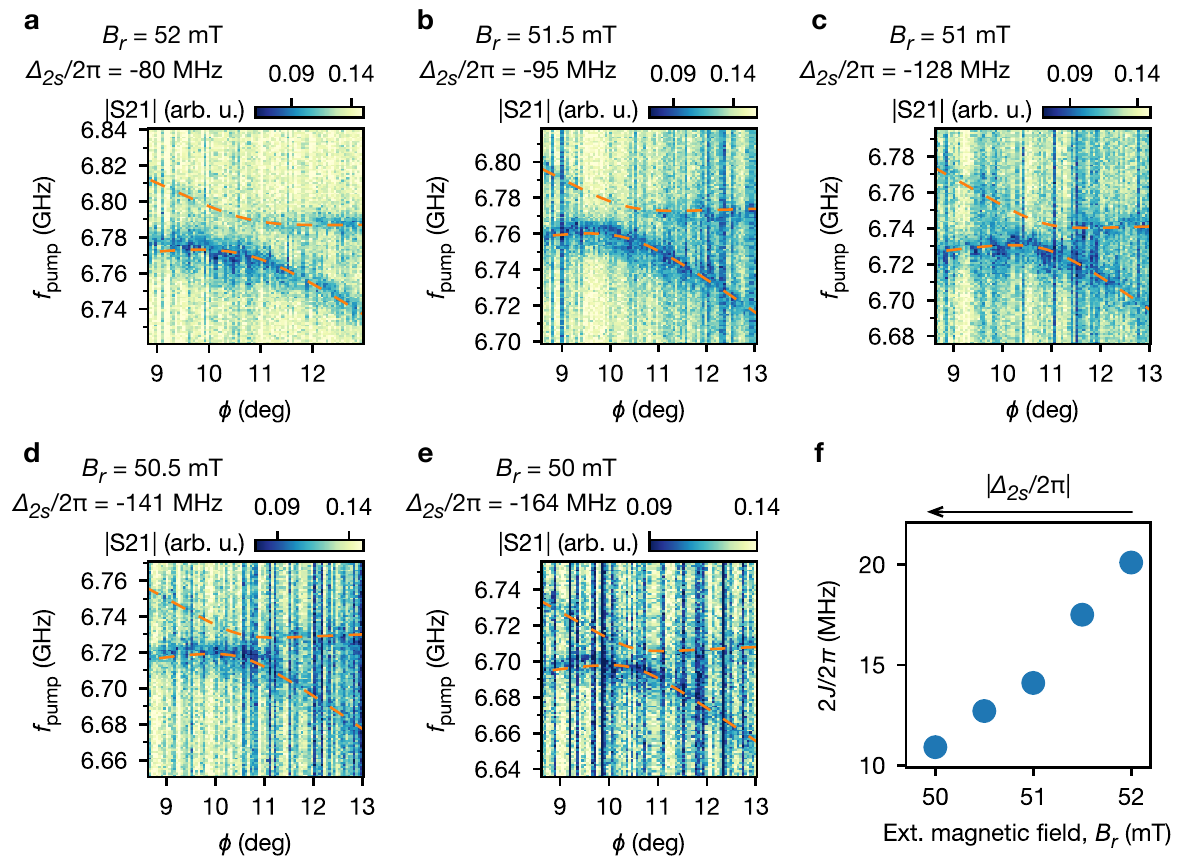} 
   \caption{{Extended spin-spin hybridization data versus spin-photon detuning.}     \figletter{a-e}~Dispersive spin-spin hybridization measured for different spin-photon detuning values. The probe power is larger for these data than for the data in \RefFigDipersivePhi{}; the shadow visible below the main transition is caused by the photon-induced dispersive shift of the spin. The apparent jitter of the transition energy as a function of $\phi$ is caused by jitter of the DQD2 tunnel coupling over the time it takes to reconstruct all angles. The $2t_c/h \approx 8.7 \GHz$ is retuned between plots, and yields $g_s/2\pi \approx 34 \MHz$ for both spins. The model fully captures the observed transition frequencies within plots and between the plots, and the only adjustment is the experimentally measured small variations in $t_c$.     \figletter{f}~Splittings $2J/2\pi$ extracted from the model as a function of spin-photon detuning (through $B_r$). Comparison with the linewidth is impaired by the photon-number broadening of the transitions. As expected, $2J/2\pi$ decreases as $|\varDelta_{2s}/2\pi|$ increases. }
   \label{fig:dispersivebr}
\end{figure*}

In this section, we present additional data that demonstrate the hybridization of the two spin states. Hybridization as a function of spin-photon detuning (through $B_r$) is shown in \figref{fig:dispersivebr}.
To speed up data acquisition, a larger probe power is used than in the main text data of \RefFigDipersivePhi{}, which, as a consequence, also significantly broadens the width of the spin transitions due to the photon-number-dependent spin dispersive shift. The apparent jitter of the transition energy as a function of $\phi$ is caused by jitter of the DQD2 tunnel coupling over the time it takes to reconstruct all angles.
As expected, the spin splitting is reduced at larger $|\varDelta_{2s}|$. The upper branch is also dark in all plots near spin-spin resonance.
These observations are consistent with spin-spin hybridization mediated by virtual resonator photons.

\begin{figure*}
   \centering
   \includegraphics{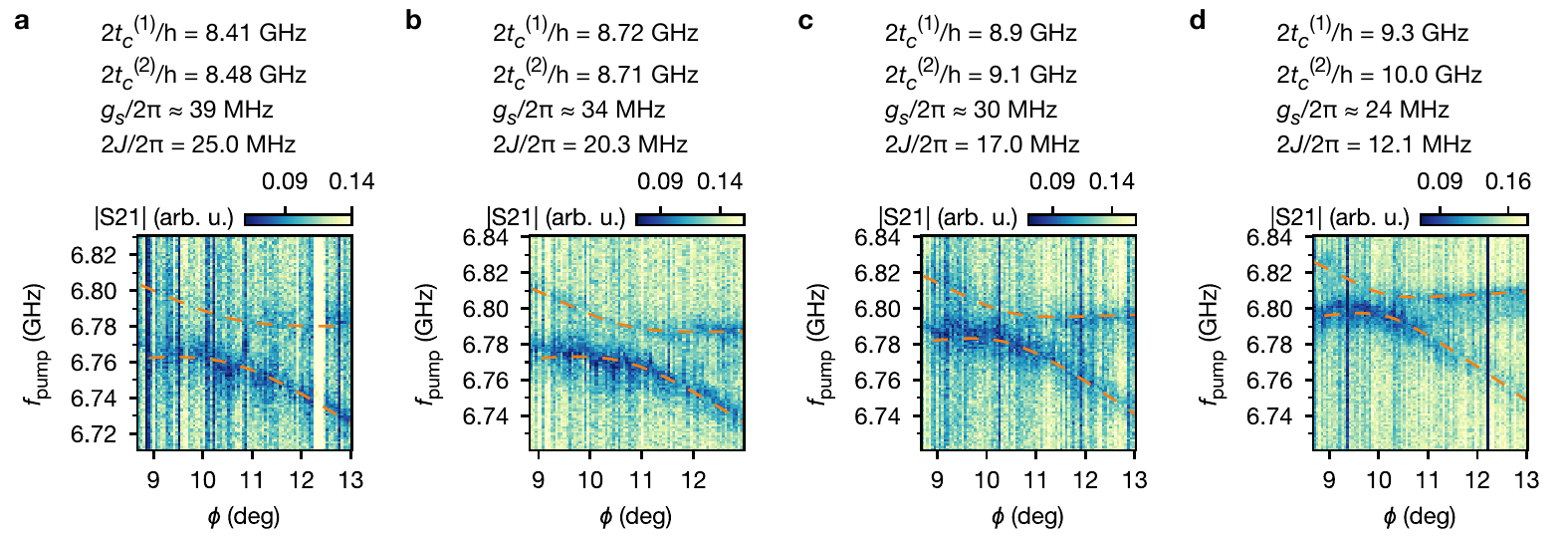} 
   \caption{{Extended spin-spin hybridization data versus spin-photon coupling strength.}     \figletter{a-d}~Dispersive spin-spin hybridization measured for decreasing spin-photon coupling values. The $g_s$ is adjusted through its dependence on $t_c$ and is approximately the same for both spins, while $B_r = 52 \mT$ is fixed. The $\varDelta_{2s}/2\pi \approx 78 \MHz$ is approximately unchanged between plots because changes in $\chi_\text{SO}$ are approximately compensated by changes in $\chi_c$. The model fully captures the observed transition frequencies within plots and between the plots, and the only adjustment is the experimentally measured $t_c$. As expected, $2J/2\pi$ decreases as $g_s/2\pi$ decreases. }
   \label{fig:dispersivetc}
\end{figure*}

Next, hybridization of the two spin states is shown as a function of the spin-photon coupling (through $t_c$) in \figref{fig:dispersivetc}.
In this work, we have measured spin-spin interactions ($J$) with values of spin-photon interactions $g_s/2\pi$ up to $40 \MHz$. A practical limit on how large $g_s$ can be is the broadening of $\kappa'_r$ at small charge-photon detunings, mainly caused by Purcell decay from the charge qubit. At some point, the resonator becomes too undercoupled and the signal too small. This could therefore be improved by using a dedicated readout resonator with an optimized coupling, by adding a near-quantum-limited amplifier at the mixing chamber, or by reducing the charge linewidth. The setting used in the main text data \RefFigDipersivePhi{} is chosen empirically based on linewidth, exchange coupling and readout signal.

In the experiment of the main text \RefFigDipersivePhi{}, the critical photon number $n_\text{crit} = (\varDelta_{2s}/2g_s)^2 = 1.5$ is quite low. This number is often used to quantify the validity of the dispersive regime, which requires the expectation value of the photon number $\bar n_r$ to be $\bar n_r \ll n_\text{crit}$ \cite{blais2021}. We use the photon-number-dependent spin dispersive shift to establish an upper bound of $\bar n_r < 0.5$ for the data of the main text \RefFigDipersivePhi{} ($\bar n_r \approx 0.4 \pm 0.1$). We can no longer distinguish between the coherent and thermal distributions since they converge and the data are too noisy. The $n_\text{crit}$ could be optimized for future qubit experiments. For instance, in \figref{fig:dispersivebr}b, the $n_\text{crit} = 2$ is already much higher while $2J/2\pi$ is still $17 \MHz$. Time-domain control can also help reduce $\bar n_r$ by probing only during readout. The exact requirements will depend on the target two-qubit gate and desired fidelity (amongst other things).

\section{Extended photon-number-dependent spin dispersive shift data}
\label{sec:photonnumber}

In this section, we extend the analysis of the photon-number-dependent spin dispersive shift data by looking at data for $\phi = 11.6 \deg$.
\begin{figure*}
   \centering
   \includegraphics{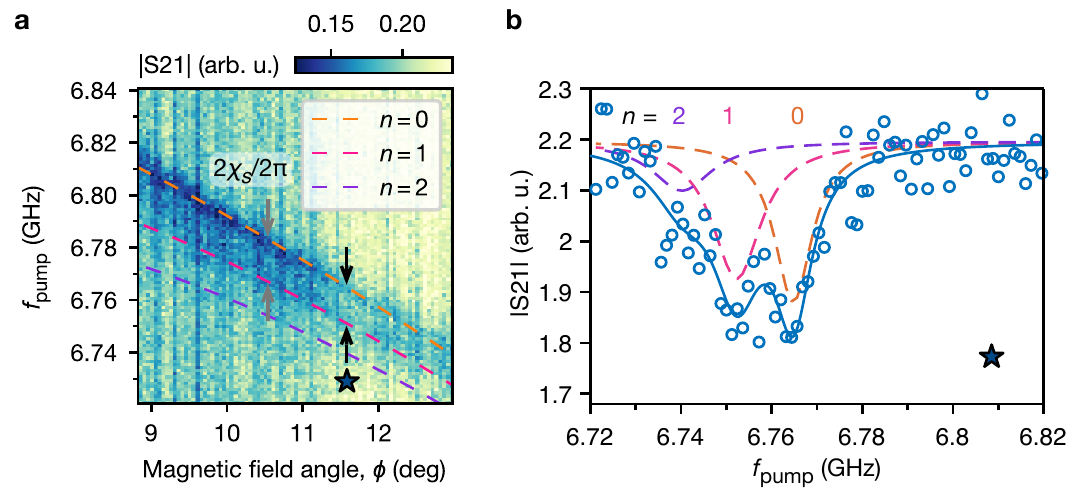} 
   \caption{{Extended photon-number-dependent spin dispersive shift data.}     \figletter{a}~Photon-number-dependent spin dispersive shift of DQD1 (same as in the main text).     \figletter{b}~Line cut of the data in panel (a) for a different angle $\phi = 11.6 \deg$, and fit to four Lorentzian dips (first three shown with dashed lines). A dispersive shift $2\chi_s/2\pi = -12.5 \MHz$ is extracted.      The coherent-state distribution has better agreement than the thermal distribution; see the main text for number distribution results.}
   \label{fig:photonnumberext}
\end{figure*}
In \figref{fig:photonnumberext}b, a line cut from \figref{fig:photonnumberext}a is extracted, and the dip areas and separations are fit to a sum of Lorentzian dips with $2\varGamma(n) = 2\gamma_s + (n + \bar n_r)\kappa'_r$. A value of $2\chi_s/2\pi = -12.5 \MHz$ is extracted from the fit ($g_s/2\pi = 33.4 \MHz$, $\varDelta_s/2\pi = -122 \MHz$), slightly larger than the average of the linewidths $(2\varGamma(0), 2\varGamma(1))/2\pi = (10.1, 13.1) \MHz$. 
The coherent state has a higher photon number, $\bar n_r = 0.82$, than at $\phi = 10.55 \deg$ (see main text, $\bar n_r = 0.62$). This can be attributed to the resonator's higher dressed quality factor for $\phi > 11.2 \deg$ (which is independently verified), and it leads to a larger probe photon population at a steady state. This is also consistent with the background showing a larger relative $|\text{S21}|$ value at fixed probe power. The difference between the coherent-state and thermal-state distributions is more pronounced than in the main text.

In \tabref{tab:params}, a small discrepancy is observed between the predicted and measured values of $\chi_s$. This could be due to driven system dynamics, which are known to modify the splitting \cite{schuster2007}.

\bibliographystyle{apsrev4-1-title} 
\bibliography{/Users/phc/Documents/Papers/PHC-ibm.bib}

\end{document}